\input epsf
\documentstyle[preprint,aps]{revtex}
\tighten
\draft
\preprint{\vbox{Submitted to Physical Review Letters \hfill IU/NTC
    96-05 \\
   \null \hfill nucl-th/9603138}}
\begin{document}

\title{Neutrino Trapping in a Supernova and Ion Screening}

\author{C. J. Horowitz~\footnote{E-mail:~Charlie@iucf.indiana.edu}}
\address{Institut fur Theoretische Kernphysik,
Universitat Bonn, D53115 Bonn,
Nussallee 14-16, Germany and \\
Nuclear Theory Center and Department of Physics, 
Indiana University,
Bloomington, Indiana 47408, USA} 
\date{\today}
\maketitle
\begin{abstract}
Neutrino-nucleus elastic scattering is reduced in dense
matter because of correlations between ions.
The static structure factor for a plasma of electrons and ions
is calculated from Monte Carlo simulations and parameterized with
a least squares fit.   Our results 
imply a large increase in the neutrino mean free path.  This strongly limits 
the trapping of neutrinos in a supernova by coherent neutral current 
interactions.  
\end{abstract}
\narrowtext

A (core collapse) supernova radiates large numbers of neutrinos.
Indeed, the energy in neutrinos is 100 times greater then
that in all other forms of matter~\cite{burrows}.  Therefore, supernova
models may depend on the details of neutrino interactions in dense
matter.  In this paper, we calculate how correlations in the medium
modify the important neutrino-nucleus elastic
scattering cross section.  This cross section is large because it
involves coherent scattering from all of the nucleons in a 
nucleus~\cite{coherent}.  However, when the neutrino wave length is
comparable to the inter particle spacing there are also coherent 
contributions from different nuclei.  These can screen
the interaction and lead to a large reduction in the cross section.

In the present supernova model, the core of a massive star runs out of 
nuclear fuel and collapses~\cite{bethe}.
This core is composed of a dense plasma of electrons and nuclei.  
As the density reaches $10^{11}$ to $10^{12}$ g/cm$^3$ the
medium starts to become opaque to neutrinos.  The neutrino opacity is
thought to be dominated by neutrino-nucleus elastic scattering (as
long as a significant fraction of the matter is in nuclei).
This opacity insures that neutrino transport involves diffusion (rather
then free streaming).  The diffusion time can become
long compared to the dynamical time scale, thus trapping neutrinos and 
their lepton number.

The neutrino-nucleus elastic cross section in free space is~\cite{cross},
\begin{equation}  
d\sigma_0/d\Omega ={G^2C^2E_\nu^2(1+{\rm cos}\theta)\over 4\pi^2},
\label{free}
\end{equation}
with $G$ the Fermi constant, $E_\nu$ the neutrino energy, $\theta$ 
the scattering angle and the
total weak charge $C$ of a nucleus of charge $Z$ and neutron number $N$ is
\begin{equation}
C=-2Z{\rm sin^2}\Theta_W +(Z-N)/2.
\end{equation}
(We use a Weinberg angle of ${\rm sin^2}\Theta_W=.223$.)  
In a dense plasma this cross section is modified by
electron~\cite{elec,ion} and ion~\cite{ion} screening.  Imagine a single
impurity ion in a dense plasma.  Extra electrons will be attracted
to the impurity.  Since these electrons 
have weak interactions they screen both the electro-magnetic and weak
charge of the ion.  However, the very dense relativistic electron gas
is quite rigid because of the large Fermi momentum.  This limits
the effect of electron screening (see below).

Other ions can also screen the impurity by creating a small hole in
the ion distribution.  At temperatures of order 
one MeV, the ions are essentially classical and their screening 
is not impeded by a large Fermi energy.  
Therefore, we will focus on ion screening in this paper.  
Some ion screening results have been presented in Ref.~\cite{ion}.
Here we calculate screening for a broad range of densities and
determine its impact on the neutrino mean free path.  We also provide
a parameterization of our results.  This, or a further simplification,
will allow the incorporation of screening in neutrino
transport codes. 

Ion screening is included by 
multiplying Eq.~(\ref{free}) by the static structure factor $S_q$ 
of the ions~\cite{noize},
\begin{equation}
d\sigma/d\Omega = d\sigma_0/d\Omega\, S_q.
\label{sigma}
\end{equation}
Here $q$ is the momentum transfer and $d\sigma/d\Omega$ the effective
cross section in the medium.  We discuss $S_q$ below.

The transport cross section is the angle integral of 
Eq.~(\ref{sigma}) with a factor of $(1-{\rm cos}\theta)$,
\begin{equation}
\sigma^t =\int d\sigma/d\Omega\, (1-{\rm cos})\, d\Omega\ =\sigma_0^t\, <S>.
\label{sigmat}
\end{equation}
The free transport cross section is,
$\sigma_0^t={2\over 3} G^2C^2E_\nu^2/\pi$,
and $<S>$ is the angle average of $S_q$,
\begin{equation}
<S>={3\over 4}\int_{-1}^1 d\,{\rm cos}\theta \, (1+\rm{ cos}\theta) 
(1-{\rm cos}\theta)\, S_{q(\theta)}.
\label{save}
\end{equation}
Here $(1+{\rm cos}\theta)$ is from the angular dependence of the free
cross section and $q(\theta)^2=2E_\nu^2(1-{\rm cos}\theta)$.  Thus, 
ion screening can be incorporated into neutrino transport codes 
be multiplying the existing interactions by the factor $<S>$
\footnote{Note,
Eqs. (\ref{free},\ref{sigmat}) have ignored axial current
contributions to the cross section.  These may be significant when
$<S>$ is very small.}.  
This depends on the density, temperature and neutrino energy.  The
transport mean free path $\lambda$ then follows,
$\lambda =1/(n\sigma^t)$,
with $n$ the number density of ions.

The static structure factor $S_q$ is determined from a Monte Carlo 
simulation~\cite{mcmil} of the radial distribution 
function $g(r)$~\cite{ich}.  This gives the probability
to find another ion a distance $r$ from a given ion and is
calculated by histograming the relative distances in the
simulation~\cite{mcmil},
\begin{equation}
S_q=1+n\int d^3r\, {\rm e}^{-i{\bf q}\cdot{\bf r}}(g(r)-1).
\label{sq}
\end{equation}
Equations~(\ref{save},\ref{sq}) yield a simple integral for $<S>$,
\begin{equation}
<S>=1+{4\pi n\over E_\nu^2}\int_0^\infty dr f(2E_\nu r)(g(r)-1),
\end{equation}
with 
\begin{equation}
f(x)=72({\rm cos}x+x{\rm sin}x-1)/x^4-6(5{\rm cos}x+x{\rm sin}x
+1)/x^2.
\end{equation}

The classical canonical partition function is 
simulated using $N_i\approx 1000$\ ions in a box of volume $V=N_i/n$ with
periodic boundary conditions.  The ions interact via screened Coulomb 
potentials,
\begin{equation}
v(r)={Z^2 e^2\over 4 \pi r}{\rm e}^{-r/\lambda_e}.
\label{screened}
\end{equation}
Here $\lambda_e=\pi/(ek_F)$ describes the electron screening of the ion-ion
interaction~\cite{fw}.  Note, this Yukawa approximation can be replaced by a
more accurate description at high momentum transfers.  However, we are
primarily interested in momentum transfers $q$ much less then the Fermi
momentum $q<<k_F$.  Therefore Eq.~(\ref{screened}) should be adequate for 
our purposes.

The system is warmed up for about 200 Metropolis sweeps starting from either
a simple cubic lattice or a uniform distribution.  
Statistics are then accumulated 
using 400 configurations each of which is 
separated by 5 sweeps.  This yields 
$S_q$ with a typical statistical accuracy of $2-3\times 10^{-3}$.  
These results are close to $S_q$\ for a pure one component 
plasma~\cite{ocp}.

We parameterize our Monte Carlo results for $<S>$ as a 
function of two dimension-less variables.  It is a strong function of
\begin{equation}
\bar E  = E_\nu a
\label{ebar}
\end{equation}
Here, the ion sphere radius $a$\ measures the average distance between
ions~\cite{ich},
\begin{equation}
a=[3/(4\pi n)]^{1/3}.
\end{equation}
Next, $<S>$ is a weak function of $\Gamma$ which characterizes the 
strength of the interaction.   This is the
ratio of a typical Coulomb potential to the thermal energy $kT$~\cite{ich},
\begin{equation}
\Gamma = {Z^2 e^2 \over 4\pi a kT},
\label{gamma}
\end{equation}
(with $e^2=4\pi\alpha\approx 0.0917$).  In general $<S>$ is a function
of the density and temperature separately.  However, if one ignores
the relatively small effect of the screening length $\lambda_e$ in
Eq. (\ref{screened}) then $<S>$ only depends on $\Gamma$ (and $\bar
E$).  We have performed simulations for a pure $^{56}$Fe plasma at
$kT=1$ MeV.  We assume results can be extrapolated to other
compositions and temperatures by calculating the appropriate $\Gamma$.

A least squares fit of our Monte Carlo results valid for all $E_\nu$
and $1<\Gamma < 150$ is carried out.  This fit is based on simulations
for twelve values of $\Gamma$ between 0.87 and 151.8.  For a temperature
of one MeV this corresponds to $^{56}$Fe densities from $2\times 10^7$ to
$9\times 10^{13}$ g/cm$^3$.  We approximate $<S>$,
\begin{equation}
<S(\bar E,\Gamma)>=1/[1+{\rm exp}( -\sum_{i=0}^6\beta_i(\Gamma) \bar E^i )],
\label{lsfit}
\end{equation}
for
\begin{equation}
\bar E < E^*(\Gamma)=3+4/\Gamma^{1/2}.
\end{equation}
While for $\bar E > E^*$ we assume,
\begin{equation}
<S(\bar E,\Gamma)>=1.
\end{equation}
The coefficient functions $\beta_i(\Gamma)$, for i=3,4,5 and 6  are expanded 
in a power series in $\Gamma^{1/2}$,
\begin{equation}
\beta_i(\Gamma) = \beta_{i1}+\beta_{i2}\Gamma^{1/2}
+\beta_{i3}\Gamma + \beta_{i4} \Gamma^{3/2}.
\end{equation}
The coefficients $\beta_{ij}$ are collected in 
table~\ref{tableone}.
Finite size effects contaminate the Monte Carlo results for small $\bar E$.
Therefore we use RPA results for $\beta_0$,
\begin{equation}
\beta_0 = {\rm ln} [0.300/(0.300 + 3\Gamma)],
\end{equation}
$\beta_1=0$ and $\beta_2 =6.667$.

The error in the fit is typically less then 0.01.
Although, for very large $\Gamma$, $<S>$ oscillates around
one at large $\bar E$.  This oscillation is not reproduced by our fit
and can lead to an error as large as 0.05.  However, this only occurs
at very high densities and is expected to have negligible impact on
the dynamics.   Again, the fit is valid for all neutrino
energies and $1 <\Gamma < 150$.  For smaller $\Gamma$ a good
estimate is provided by simply setting $\Gamma=1$.  (Note, here
$<S>$\ is only important at very small neutrino energies.)
Likewise, for $\Gamma> 150$\ a reasonable estimate is provided by setting
$\Gamma=150$ (as long as the system is in the liquid phase).  A solid
is expected to form for $\Gamma \approx 180$\cite{solid1}.  This may be
relevant for models of type Ia supernovae\cite{solid2}.  The very interesting
problem of ``Bragg diffraction'' of neutrinos in a radioactive crystal
remains to be investigated.  Neutrino wave lengths can be
comparable to the lattice spacing.

We use this fit for $<S>$ to calculate the mean free path of a
neutrino in a plasma of ions, neutrons and electrons.
For example, Cooperstein and Wambach~\cite{wambach} modeled 
matter at $10^{12}$  g/cm$^3$ as consisting of $X_n=6$ percent free 
neutrons and 94 percent nuclei of average
charge $Z\approx 37$\ and average mass $A\approx 97$ at a temperature
of 1.5 MeV.  This is appropriate for the collapse phase of a 
supernova.  We use this composition in calculating the mean free
path.  For simplicity, the composition and
temperature are assumed not to change with density and we ignore the strong
interactions between ions and or neutrons.

The transport mean free path $\lambda$ is assumed dominated by elastic 
scattering off of nuclei and neutrons~\cite{brown}
\begin{equation}
\lambda = {{15\ \rm km}\over \rho_{12}} 
\Bigr({{10\ \rm MeV}\over E_\nu}\Bigl)^2 [(1-X_n){C^2\over A}<S>R_e
+ X_n({c_v^n}^2+5{c_a^n}^2)]^{-1}.
\label{mfp}
\end{equation}
Here $\rho_{12}$ is the density in units of $10^{12}$ g/cm$^3$, the
weak couplings of a neutron $c^n_i$ are given in ref.~\cite{wer} and
$R_e$\ is an additional correction factor that describes electron
screening.  This is calculated in ref.~\cite{ion} and can be
approximated, see below Eq.~(\ref{screened}),
\begin{equation}
R_e\approx \Biggl[1+\bigl({c_v^eZ\over C}\bigr)
{1\over 1 + 2.5E_\nu^2\lambda_e^2}\Biggr]^2.
\label{elec}
\end{equation}
Each ion has an electron cloud around it.  Electron neutrinos or
anti-neutrinos couple to this with strength,
$c_v^e=2{\rm sin}^2\Theta_W+{1\over 2}$,
while muon neutrinos do not see the electron cloud
$c_v^e\approx 0$ and thus $R_e\approx 1$.

The mean free path $\lambda$ is shown in Fig. 1.
To our knowledge, almost all present supernova simulations use
Eq.~(\ref{mfp}) with $<S>=R_e=1$.  This leads to a very small
mean free path (which traps neutrinos for densities of about
$\rho_{12}=0.5$ and above).  However, including $<S>$ leads to
a dramatic increase in $\lambda$ and to a large change in its
density dependence.   The rapid decrease of $<S>$ with density 
can lead to a $\lambda$ which actually {\it increases} with 
density.  
Over a range of densities $\lambda$ for $E_\nu=10$ MeV is 
greater then 10 km.  
This is much larger then the unscreened $\lambda$ 
($\approx 0.4$ km at $\rho_{12}=5$).  Finally,
electron screening causes $\lambda$ for a $\nu_e$ to be about 15 
percent {\it larger} then for a $\nu_\mu$.

Screening effects are even more important for lower $E_\nu$.
For example at 5 MeV, $\lambda$ is greater then 45 km even at
$\rho_{12}=10$.
This is larger then the size of the dense system ($\approx$ 30 km) 
so a neutrino-sphere may not form at all (for this energy).

Figure 1 shows that $\lambda$ {\it is larger then the size of 
the system for $E_\nu$ less then or equal to about 7.5 MeV}.  
For $E_\nu$ between 7.5 and about 10 MeV the relatively large 
$\lambda$ will allow neutrinos to diffuse out of the system (in 
about a msec or less).  These are main results of this paper.

However, at $E_\nu=20$ MeV (or above)
screening is reduced and the overall $1/E^2$ scale of $\lambda$
is smaller so that the mean free path is significantly 
shorter.
Screening is not very sensitive to temperature (as long as 
there are no large changes in composition).  Changing $T$ 
leads to a change in $\Gamma$\, see Eq.~(\ref{gamma}).  However  
$<S>$ is not a strong function of $\Gamma$.  Likewise $<S>$ is 
not very sensitive to the average Z of the material.  Changes 
in the average A change $a$ in Eq.~(\ref{ebar}) and the 
overall factor $C^2/A$ in Eq.~(\ref{mfp}).  Thus $\lambda$ 
decreases with increasing A.

The static structure factor $S_q$ describes the total strength to
scatter from the medium.   At high density, this may be 
dominated by the excitation of {\it ion} plasma oscillations rather
then (quasi-) elastic scattering.  These plasma osc. have
energies $\omega\approx (Z^2e^2n/M_i)^{1/2}$ (with ion mass $M_i$)
and may lead to a much larger net neutrino energy loss per $\nu-A$
collision~\cite{tobepub}.  This can impact neutrino thermalization and
heating.

Screening effects will be all but absent after the supernova
shock wave dissociates the nuclei.  Then $\lambda$\ will be relatively 
short because of scattering from large numbers of nearly free neutrons
and protons.  Thus, {\it the neutrino opacity is small (because of screening)
before the shock wave and large afterwards.}  This change in opacity
could have important effects on the dynamics.  Perhaps the situation
is not unlike the photon opacity of the universe being large before
and small after recombination.

We now speculate on some of the implications of screening on supernova
simulations.  We emphasize that final conclusions await neutrino transport 
calculations.   First, more neutrinos may escape allowing
additional electron capture  (as escape reduces
the neutrino chemical potential).   Electron capture reduces the pressure 
and the energy of the shock.  

For example a 20 MeV (or higher) electron capture neutrino produced at 
$\rho_{12}=5$ can have its energy reduced to about 10 MeV in 
a time of order 1/3 msec by electron scattering.
The neutrino can then diffuse out of the core in 
about a msec.  Alternatively the neutrino's energy can be reduced to
7.5 MeV (in about a msec) and then directly escape.

Second, increased diffusion should raise the neutrino luminosity during
the early stages of a supernova.  
[Perhaps, this could be observable
if many prompt $\nu_e$s are detected from a nearby supernova~\cite{burrows}.]
This may enhance the neutrino transport of energy to the shock.  
Screening has almost no effect on the opacity of low density matter or
of the dissociated material after shock
passage.  Thus screening should not interfere with the ability of material
near the shock to absorb energy from neutrinos.   (Note, screening
introduces a new density dependence to the neutrino interactions.)

We have calculated the effect of ion screening on neutrino-nucleus elastic
scattering.  Our Monte Carlo results for the angle average of the static
structure factor have been fitted to an analytic formula.  This may allow
the inclusion of screening in simulations.  We find that
the mean free path of a 10 MeV (or lower) neutrino is greatly increased. 
This could have important effects on the early stages of a supernova.

We thank the Institut fur Theoretische Kernphysik for their kind
hospitality. This research was supported in part by the US Department
of Energy under Grant No. DE-FG02-87ER-40365 and by the Deutsche 
Forschungsgemeinschaft.

\mediumtext
\begin{table}
\caption{Parameters $\beta_{ij}$ from a least squares fit
  of the angle averaged static structure factor $<S>$, see text.}
\begin{tabular} {ccccc}
Coeff. & $j=1$ & 2 & 3 & 4 \\
\tableline
$\beta_{3j}$ & -7.362056 & 0.5371365 & -0.1078845 & 4.189612E-3 \\
$\beta_{4j}$ & 3.4489581 & -0.40251656 & 9.0877878E-2 & -3.4353581E-3 \\
$\beta_{5j}$ & -0.74128645 & 0.11019855 & -2.5359361E-2 & 9.0487744E-4 \\
$\beta_{6j}$  & 5.9573285E-2 & -1.0186552E-2  & 2.2791369E-3 & -7.4614597E-5 \\
\end{tabular}
\label{tableone}
\end{table}


\begin{figure}
\epsfbox{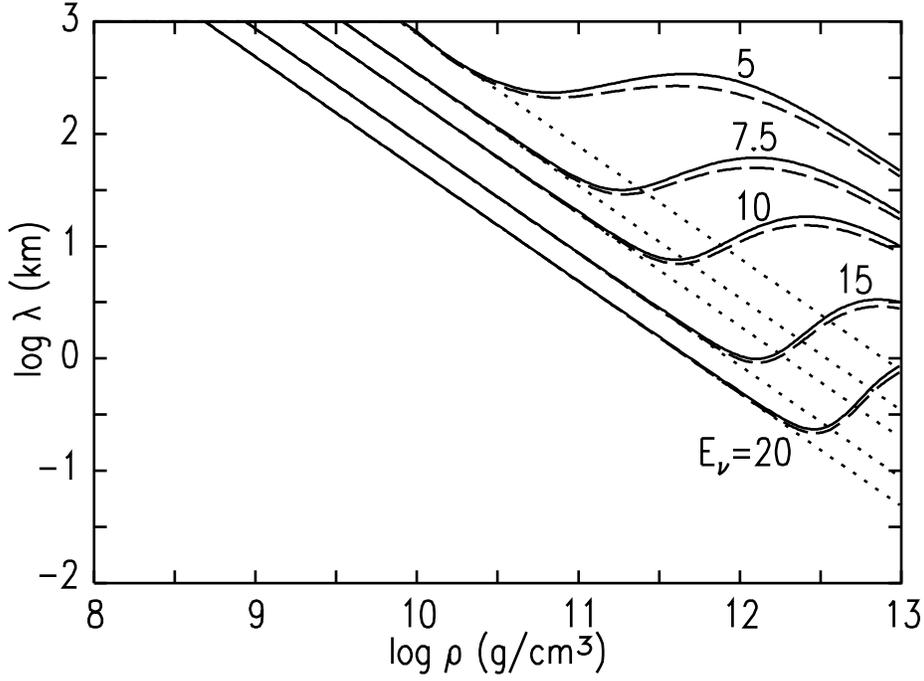}
\nobreak
\caption{Neutrino transport mean free path vs. density.  The solid
  lines include both ion $<S>\neq 1$ and electron $R_e\neq 1$
  screening and are appropriate for $\nu_e$, $\bar \nu_e$ while the dashed
  lines for $\nu_\mu$ neglect electron screening $R_e=1$.  Finally the
  dotted lines (used in most present supernova simulations) neglect
  all screening $<S>=R_e=1$.  
Top to bottom, the curves are for neutrino energies of $E_\nu=5$, 7.5, 10, 15, 
and 20 MeV.}
\label{logmfpfig}
\end{figure}

\end{document}